\documentstyle[11pt]{article}
\begin{document}


\title{ 
{\bf Supersymmetry and Singular Potentials}}
\author{Ashok Das \\
\\
Department of Physics and Astronomy, \\
University of Rochester,\\
Rochester, New York, 14627\\
\\
and\\
\\
Sergio A. Pernice\\
\\
Universidad del CEMA\\
Av. C\'{o}rdoba 374\\
1054 Capital Federal, Argentina. }
\date{}
\maketitle

\begin{abstract}
The breaking of supersymmetry due to singular potentials
in supersymmetric quantum mechanics is critically analyzed. It
is shown that, when properly regularized, these potentials
respect supersymmetry, even when the regularization parameter is removed. 
\end{abstract}

\vfill\eject
\section{Introduction}

Supersymmetry is a beautiful and, simultaneously, a tantalizing
symmetry [1-7]. On the one hand, supersymmetry leads to field theories and
string theories with exceptional properties [8-9]. The improved ultraviolet
behavior, the natural solution of the hierarchy problem are just a few
of the nice features of supersymmetric theories. On the other hand,
supersymmetry also predicts degenerate superpartner states (superpartner
states with degenerate mass) corresponding to every physical particle
state of the theory. Unfortunately, such superpartners with degenerate
masses are not observed experimentally and, consequently, one expects
that supersymmetry must be spontaneously (dynamically) broken much
like the spontaneous breaking of ordinary symmetries in physical
theories. However, unlike ordinary symmetries, spontaneous breaking of
supersymmetry has so far proved extremely difficult in the
conventional framework. Consequently, in the context of supersymmetry,
one constantly looks for alternate, unconventional methods of breaking
of this symmetry [6-7]. There is, of course, the breaking of supersymmetry
due to instanton effects which is well understood. However, several
authors, in recent years have suggested that supersymmetry may be
broken in the presence of singular potentials or boundaries in a
nonstandard manner [10-12]. We will explain the details of the proposed
mechanism later, but the gist of the argument is that in such systems,
the superpartner states may not belong to the physical Hilbert space
thereby leading to a breaking of supersymmetry. This would, of course,
explain why the superpartner states would not be observable. Even more
interesting is the possibility that since such a breaking
is nonstandard, the usual theorems of supersymmetry breaking may not
apply and the ground state may continue to have zero energy
thereby leading to a solution of the cosmological constant problem also.

The examples, where such a breaking has been discussed, are simple
quantum mechanical models which nonetheless arise from the
non-relativistic limit of some field theories. It is for this reason
that, in an earlier paper, we had examined [13] a candidate relativistic
$2+1$ dimensional field theory to see if the manifestation of such a
mechanism was possible in a field theory. However, a careful
examination  of the theory revealed that
supersymmetry prevails at the end although  it might appear naively, in
the beginning, that  supersymmetry would be broken in the nonstandard
manner. This prompted us to re-analyze the quantum mechanical models,
where this mechanism was demonstrated, more carefully in order to
understand if supersymmetry is truly broken in these models and if so
what may be the distinguishing features in these models from a
relativistic field theory. A  systematic and critical examination,
once again, reveals that when carefully done, supersymmetry is
manifest even in such singular quantum mechanical models which is the
main result of this paper.

Since our discussion would be entirely within the context of one
dimensional supersymmetric quantum mechanics, let us establish the
essential notations here. Given a superpotential, $W(x)$, we can
define a pair of supersymmetric potentials as
\begin{equation}
 V_{+} = {1\over 2}\left(W^{2}(x) + W'(x)\right),\qquad V_{-} =
 {1\over 2}\left(W^{2}(x) - W'(x)\right)\label{a1}
\end{equation}
where \lq\lq prime'' denotes differentiation with respect to $x$. With
$\hbar =1$ and $m=1$, we can, then, define a pair of Hamiltonians
which describe a supersymmetric system as
\begin{eqnarray}
H_{+} & = & - {1\over 2}\frac{d^{2}}{dx^{2}} + V_{+}\nonumber\\
H_{-} & = & - {1\over 2}\frac{d^{2}}{dx^{2}} + V_{-}\label{a2}
\end{eqnarray}
In fact, defining the supercharges as
\begin{equation}
Q = {1\over\sqrt{2}}\left(-\frac{d}{dx} + W(x)\right), \qquad
Q^{\dagger} = {1\over\sqrt{2}}\left(\frac{d}{dx} + W(x)\right)\label{a3}
\end{equation}
we recognize that we can write the pair of Hamiltonians in
eq. (\ref{a2}) also as
\begin{equation}
H_{+} = Q^{\dagger}Q, \qquad H_{-} = QQ^{\dagger}\label{a4}
\end{equation}

It is clear now that if $|\psi_{-}\rangle$ represents an eigenstate of
$H_{-}$ with a nonzero energy $E$, then, $Q^{\dagger}|\psi_{-}\rangle$
would be an eigenstate of $H_{+}$ with the same energy, namely,
\begin{eqnarray}
H_{-}|\psi_{-}\rangle & = & QQ^{\dagger}|\psi_{-}\rangle =
E |\psi_{-}\rangle\nonumber\\
H_{+}(Q^{\dagger}|\psi_{-}\rangle) = Q^{\dagger}Q(Q^{\dagger}|\psi_{-}\rangle)
& = & Q^{\dagger}(QQ^{\dagger}|\psi_{-}\rangle) = E
(Q^{\dagger}|\psi_{-}\rangle)\label{a5} 
\end{eqnarray}
In other words, the states $|\psi_{-}\rangle$ and
$Q^{\dagger}|\psi_{-}\rangle$ (alternately, we can also denote them as
$Q|\psi_{+}\rangle$  and $|\psi_{+}\rangle$ respectively)
would correspond to the degenerate superpartner states. All the
eigenstates of the two Hamiltonians $H_{+}$ and $H_{-}$ would be
degenerate except for the ground state with vanishing energy which
would correspond to the state satisfying
\begin{equation}
Q|\psi_{+}\rangle = 0, \qquad {\rm or,}\qquad Q^{\dagger}|\psi_{-}\rangle =
0\label{a6} 
\end{equation}

For a given superpotential, at  most one of the two conditions in
eq. (\ref{a6}) can be satisfied (that is, at  most, only one of the
two conditions in (\ref{a6}) would give a normalizable state). Namely,
the ground state with vanishing energy is unpaired and can belong to
the spectrum of either $H_{+}$ or $H_{-}$ depending on which of the
conditions leads to a normalizable state. This corresponds to the case
of unbroken supersymmetry. For example, with $W(x)=-\omega x$, the pair
of Hamiltonians
\begin{eqnarray}
H_{+} & = & {1\over 2}\left(-\frac{d^{2}}{dx^{2}} + \omega^{2}x^{2} -
\omega\right)\nonumber\\
H_{-} & = & {1\over 2}\left(-\frac{d^{2}}{dx^{2}} + \omega^{2}x^{2} +
\omega\right)\label{a7}
\end{eqnarray}
describe the supersymmetric harmonic oscillator and it can be easily
checked that the normalizable ground state belongs to the spectrum of
$H_{+}$ (In fact, we will choose $H_{+}$ to have the ground state
throughout this paper.). 

If, on the other hand, the superpotential is such that neither of the
states  in eq. (\ref{a6}) is normalizable, then, supersymmetry is
known  to be broken by instanton effects [6]. For example, this can happen
when the superpotential is an even polynomial. In contrast, the new
mechanism described, in the presence of a boundary or a singular
potential [10-12], corresponds to the case where the action of the
supercharges takes a state out of the physical Hilbert space (although
$QQ^{\dagger}$ and $Q^{\dagger}Q$ belong to the Hilbert space) so that,
say, if $|\psi_{+}\rangle$ is in the Hilbert space,
$Q|\psi_{+}\rangle$ would not belong to the Hilbert space leading to
the breaking of the degeneracy of states and, therefore, supersymmetry.

In this paper, we carefully analyze the models [10-12] where this mechanism
is thought to be operative. We note that a quantum mechanical
potential with a singular structure is best studied with a
regularization because this  brings out the correct boundary conditions
naturally. Furthermore, the regularization must be chosen carefully
preserving supersymmetry when one is dealing with a supersymmetric
potential with a singular structure. Keeping this in mind, we examine a
supersymmetric model with a boundary, namely, the supersymmetric
oscillator on the half line [10] in section {\bf 2}. We briefly
recapitulate the  results of a harmonic 
oscillator  on the half line and then show through a careful analysis
that supersymmetry is, in fact, manifest in such a model in spite of
the boundary. In section {\bf 3}, we analyze a model with a singular
potential, namely, the oscillator with a ${1\over x^{2}}$
potential. This potential is quite interesting and, in spite of
several studies in the literature [14-16], the analysis is not quite
complete and we present a systematic and complete analysis of this
problem. In section {\bf 4}, we take up the study of the
supersymmetric oscillator with a ${1\over x^{2}}$ potential [11] and show
that supersymmetry is, in fact, manifest in this system as well, in
spite of the singular nature of the potential. There
are several interesting aspects of this model which emerge from a
careful analysis which we bring out. In section {\bf 5}, we solve this problem
algebraically as well which supports the analysis of section {\bf
4}. In section {\bf 6}, we describe the solution to a puzzle raised in
the literature [11] in the context of the supersymmetric oscillator with a
${1\over x^{2}}$ potential and present our conclusions in section {\bf 7}.  

\section{Super \lq\lq Half'' Oscillator}

To understand the super \lq\lq half'' oscillator, it is useful to
recapitulate briefly the results of the \lq\lq half'' oscillator. Let
us consider a particle moving in the potential
\begin{equation}
V(x) = \left\{\begin{array}{lll}
               {1\over 2}(\omega^{2}x^{2}-\omega)& {\rm for}& x>0\\
                \infty & {\rm for} & x<0
              \end{array}\right.\label{b1}
\end{equation}
The spectrum of this potential is quite clear intuitively. Namely,
because of the infinite barrier, we expect the wave function to vanish at the
origin leading to the conclusion that, of all the solutions of the
oscillator on the full line, only the odd solutions (of course, on
the \lq\lq half'' line there is no notion of even and odd) would survive in
this case. While this is quite obvious, let us analyze the problem
systematically for later purpose.

First, let us note that singular potentials are best studied in a
regularized manner because this is the only way that  appropriate
boundary conditions can be determined correctly. Therefore, let us
consider the particle moving in the
regularized potential
\begin{equation}
V(x) = \left\{\begin{array}{lll}
               {1\over 2}(\omega^{2}x^{2}-\omega)& {\rm for}& x>0\\
                  &   &\\
               {c^{2}\over 2} & {\rm for} & x<0
              \end{array}\right.\label{b2}
\end{equation}
with the understanding that the limit $|c|\rightarrow\infty$ is to be
taken at the end. The Schr\"{o}dinger equation
\[
\left(-{1\over 2}\frac{d^{2}}{dx^{2}} + V(x)\right)\psi(x) = \epsilon
\psi(x) 
\]
can now be solved in the two regions. Since $|c|\rightarrow\infty$ at
the end, for any finite energy solution, we have the asymptotically
damped solution, for $x<0$,
\begin{equation}
\psi^{(II)}(x) = A\,e^{(c^{2} - 2\epsilon)^{{1\over 2}}\,x}\label{b3}
\end{equation}

Since the system no longer has reflection symmetry, the solutions, in
the region $x>0$,
cannot be classified into even and odd solutions. Rather, the
normalizable (physical)
solution would correspond to one which vanishes asymptotically. The
solutions of the Schr\"{o}dinger equation, in the region $x>0$, are
known as the parabolic cylinder functions [17] and the asymptotically
damped physical solution is given by
\begin{equation}
\psi^{(I)}(x) = B\,U(-({\epsilon\over\omega}+{1\over 2}),
\sqrt{2\omega}\,x)\label{b4} 
\end{equation}
The parabolic cylinder function, $U(a,x)$, of course, vanishes for
large values of $x$. For small values of $x$, it satisfies
\begin{eqnarray}
U(a,x) &\stackrel{x\rightarrow 0}{\longrightarrow}&
\frac{\sqrt{\pi}}{2^{{1\over 4}(2a+1)}\Gamma({3\over 4}+{a\over
2})}\nonumber\\
U'(a,x) &\stackrel{x\rightarrow 0}{\longrightarrow}&
-\frac{\sqrt{\pi}}{2^{{1\over 4}(2a-1)}\Gamma({1\over
4}+{a\over2})}\label{b5}
\end{eqnarray}
It is now straightforward to match the solutions in eqs. (\ref{b3},
\ref{b4}) and their first derivatives across the boundary at $x=0$ and
their ratio gives
\begin{equation}
\frac{1}{\sqrt{c^{2}-2\epsilon}} =
-\frac{1}{2\sqrt{\omega}}\frac{\Gamma(-{\epsilon\over
2\omega})}{\Gamma(-{\epsilon\over 2\omega}+{1\over 2})}\label{b6}
\end{equation}
It is clear, then, that as $|c|\rightarrow \infty$, this can be
satisfied only if
\begin{equation}
-{\epsilon\over 2\omega} + {1\over 2}\;
 \stackrel{|c|\rightarrow\infty}{\longrightarrow}\; -n,\qquad\qquad
 n=0,1,2,\cdots\label{b7} 
\end{equation}
In other words, when the regularization is removed, the energy levels
that survive are the odd ones, namely, (remember that the zero point
energy is already subtracted out in (\ref{b1}) or (\ref{b2}))
\begin{equation}
\epsilon_{n} = \omega (2n + 1)\label{b8}
\end{equation}
The corresponding physical wave functions are nontrivial only on the half
line $x>0$ and have the form
\begin{equation}
\psi_{n}(x) = B_{n}\,U(-(2n+{3\over 2}), \sqrt{2\omega}\,x) =
\tilde{B}_{n}\,e^{-{1\over 2}\omega
x^{2}}\,H_{2n+1}(\sqrt{\omega}\,x)\label{b9}
\end{equation}
Namely, only the odd Hermite polynomials survive leading to the fact
that the wave function vanishes at $x=0$. Thus, we see that the
correct boundary condition naturally arises from regularizing the
singular potential and studying the problem systematically.

We now turn to the analysis of the supersymmetric oscillator on
the half line. One can define a superpotential [10]
\begin{equation}
W(x) = \left\{\begin{array}{cll}
              -\omega x & {\rm for} & x>0\\
              \infty & {\rm for} & x<0
              \end{array}\right.
\end{equation}
which would, naively, lead to the pair of potentials
\begin{equation}
V_{\pm}(x) = \left\{\begin{array}{cll}
                    {1\over 2}(\omega^{2} x^{2} \mp \omega) & {\rm
                    for} & x>0\\
                    \infty & {\rm for} & x<0
                    \end{array}\right.
\end{equation}
Since, this involves  singular potentials, we can study it, as before,
by  regularizing the singular potentials as 
\begin{eqnarray}
V_{+}(x) & = &  \left\{\begin{array}{lll}
               {1\over 2}(\omega^{2}x^{2}-\omega)& {\rm for}& x>0\\
                  &   &\\
               {c_{+}^{2}\over 2} & {\rm for} & x<0
              \end{array}\right.\nonumber\\
V_{-}(x) & = &  \left\{\begin{array}{lll}
               {1\over 2}(\omega^{2}x^{2}+\omega)& {\rm for}& x>0\\
                  &   &\\
               {c_{-}^{2}\over 2} & {\rm for} & x<0
              \end{array}\right.\label{b10}
\end{eqnarray}
with the understanding that $|c_{\pm}|\rightarrow\infty$ at the end.

The earlier analysis can now be repeated for the pair of potentials in
eq. (\ref{b10}). It
is straightforward and without going into details, let us simply note
the results, namely, that, in this case, we obtain
\begin{eqnarray}
\epsilon_{+,n} & = & \omega(2n+1)\qquad
\psi_{+,n}(x)=B_{+,n}\,e^{-{1\over 2}\omega
x^{2}}\,H_{2n+1}(\sqrt{\omega}\,x)\nonumber\\
\epsilon_{-,n} & = & 2\omega(n+1)\qquad
\psi_{-,n}(x)=B_{-,n}\,e^{-{1\over 2}\omega
x^{2}}\,H_{2n+1}(\sqrt{\omega}\,x)\label{b11}
\end{eqnarray}
Here $n=0,1,2,\cdots$. There are several things to note from this
analysis. First, only the odd Hermite polynomials survive as physical
solutions since the wave function has to vanish at the origin. This
boundary condition
arises from a systematic study involving a regularized
potential. Second, the energy levels for the supersymmetric pair of
Hamiltonians are no longer degenerate. Furthermore, the state with
$\epsilon = 0$ no longer belongs to the Hilbert space (since it
corresponds to an even Hermite polynomial solution). This leads to the
conventional conclusion that supersymmetry is broken in such a case
and let us note, in particular, that in such a case, it would appear
that the superpartner states do not belong to the physical Hilbert
space (Namely, in this case, the supercharge is an odd operator and
hence connects even and odd Hermite polynomials. However, the boundary
condition selects out only odd Hermite polynomials as belonging to the
physical Hilbert space.).

There is absolutely no doubt that supersymmetry is broken in this
case. The question that needs to be addressed is whether it is a
dynamical property of the system or an artifact of the regularization
(and, hence the boundary condition) used. The answer is quite obvious,
namely, that
supersymmetry is broken mainly because the regularization (and,
therefore, the boundary condition) breaks
supersymmetry. In other words, for any value of the regularizing
parameters, $c_{\pm}$ (even if $|c_{+}|=|c_{-}|$), the pair of
potentials in eq. (\ref{b10}) do not define a supersymmetric system
and hence the regularization itself breaks
supersymmetry. Consequently, the breaking of supersymmetry that
results when the regularization is removed cannot be trusted as a
dynamical effect.

\subsection*{Regularized Superpotential}

Another way to understand this is to note that for a supersymmetric system,
it is not the potential that is fundamental. Rather, it is the
superpotential which gives the pair of supersymmetric potentials
through  Riccati type relations. It is natural, therefore, to
regularize the superpotential which would automatically lead to a pair of
regularized potentials which would be supersymmetric for any value of
the regularization parameter. Namely, such a regularization will
respect supersymmetry and, with such a regularization, it is, then, 
meaningful to ask if supersymmetry is broken when the regularization
parameter is removed at the end. With this in mind, let us look at the
regularized superpotential
\begin{equation}
W(x) = -\omega x\theta(x) + c\theta(-x)\label{b12}
\end{equation}
Here $c$ is the regularization parameter and we are supposed to take
$|c|\rightarrow\infty$ at the end. Note that although, at this level,
both the signs of the regularization parameter are allowed, existence
of a normalizable ground state (see eq. (\ref{a6})) selects out $c>0$
(otherwise, the regularization would have broken supersymmetry through
instanton effects as we have mentioned earlier).

The regularized superpotential now leads to the pair of regularized
supersymmetric potentials
\begin{eqnarray}
V_{+}(x) & = & {1\over 2}\left[(\omega^{2}x^{2}-\omega)\theta(x) +
c^{2}\theta(-x) - c\delta(x)\right]\nonumber\\
V_{-}(x) & = & {1\over 2}\left[(\omega^{2}x^{2}+\omega)\theta(x) +
c^{2}\theta(-x) + c\delta(x)\right]\label{b13}
\end{eqnarray}
which are supersymmetric for any $c>0$. Let us note that the
difference here from the earlier case where the potentials were
directly regularized (see eq. (\ref{b10})) lies only in the presence
of the $\delta(x)$ 
terms in the potentials. Consequently, the earlier solutions in the
regions $x>0$ and $x<0$ continue to hold. However, the matching
conditions are now different because of the delta function terms. 
Carefully matching the wave function and the discontinuity of the
first derivative across $x=0$ for each of the wavefunctions and taking
their ratio, we obtain the two conditions
\begin{eqnarray}
\frac{1}{(c^{2}-2\epsilon_{+})^{1/2} - c} & = & -{1\over
2\sqrt{\omega}}\frac{\Gamma(-{\epsilon_{+}\over
2\omega})}{\Gamma(-{\epsilon_{+}\over 2\omega}+{1\over
2})}\label{b14}\\
  &  & \nonumber\\
\frac{1}{(c^{2}-2\epsilon_{-})^{1/2} + c} & = & -{1\over
2\sqrt{\omega}}\frac{\Gamma(-{\epsilon_{-}\over
2\omega}+{1\over 2})}{\Gamma(-{\epsilon_{-}\over
2\omega}+1)}\label{b15}
\end{eqnarray}
It is now clear that, as $c\rightarrow\infty$, (\ref{b14}) and
(\ref{b15}) give respectively
\begin{eqnarray}
\epsilon_{+,n} & = & 2\omega n\nonumber\\
  &  &  \qquad\hspace{.5in} n=0,1,2,\cdots\nonumber\\
\epsilon_{-,n} & = & 2\omega(n+1)\label{b16}
\end{eqnarray}
The corresponding wave functions, in this case, have the forms
\begin{eqnarray}
\psi_{+,n}(x) & = & B_{+,n}\,e^{-{1\over 2}\omega
x^{2}}\,H_{2n}(\sqrt{\omega}\,x)\nonumber\\
\psi_{-,n}(x) & = & B_{-,n}\,e^{-{1\over 2}\omega
x^{2}}\,H_{2n+1}(\sqrt{\omega}\,x)\label{b17}
\end{eqnarray}
This is indeed quite interesting for it shows that the spectrum of
$H_{+}$ contains the ground state with vanishing energy. Furthermore,
all the other states of $H_{+}$ and $H_{-}$ are degenerate in energy
corresponding to even and odd Hermite polynomials as one would expect
from superpartner states. Consequently, it is quite clear that if the
supersymmetric \lq\lq half'' oscillator is defined carefully by
regularizing the superpotential, then, supersymmetry is manifest in
the limit of removing the regularization. This should be  contrasted
with  the general belief that supersymmetry is broken in this system
(which is a consequence of using boundary conditions or, equivalently,
of regularizing the potentials in a manner which violates supersymmetry).

\subsection*{Alternate Regularization}

Of course, we should worry at this point as to how regularization
independent our conclusion really is. Namely, our results appear to
follow from the matching conditions in the presence of singular delta
potential terms and, consequently, it is worth investigating whether
our conclusions would continue to hold with an alternate
regularization of the superpotential which would not introduce such
singular terms to the potentials. With this in mind, let us choose a
regularized superpotential of the form
\begin{equation}
W(x) = -\omega x \theta(x) - \lambda x \theta(-x)\label{b18}
\end{equation}
Here $\lambda$ is the regularization parameter and we are to take the
limit $|\lambda|\rightarrow\infty$ at the end. Once again, we note
that, although both signs of $\lambda$ appear to be allowed, existence
of a normalizable ground state (see eq. (\ref{a6})) would select
$\lambda>0$.

This regularized superpotential would now lead to the pair of
supersymmetric potentials of the form
\begin{eqnarray}
V_{+}(x) & = & {1\over 2}\left[(\omega^{2}x^{2}-\omega)\theta(x) +
(\lambda^{2}x^{2}-\lambda)\theta(-x)\right]\nonumber\\
V_{-}(x) & = & {1\over 2}\left[(\omega^{2}x^{2}+\omega)\theta(x) +
(\lambda^{2}x^{2}+\lambda)\theta(-x)\right]\label{b19}
\end{eqnarray}
There are no singular delta potential terms with this
regularization. In fact, the regularization merely introduces a
supersymmetric pair of oscillators for $x<0$ whose frequency is to be
taken to infinity at the end.

Since there is a harmonic oscillator potential for both $x>0$ and
$x<0$, the solutions are straightforward. They are the parabolic
cylinder functions which we have mentioned earlier. Now matching the
wave function and its first derivative at $x=0$ for each of the
Hamiltonians and taking the ratio, we obtain
\begin{eqnarray}
{1\over \sqrt{\lambda}}\frac{\Gamma(-{\epsilon_{+}\over
2\lambda})}{\Gamma(-{\epsilon_{+}\over 2\lambda}+{1\over 2})} & = &
{1\over \sqrt{\omega}}\frac{\Gamma(-{\epsilon_{+}\over
2\omega})}{\Gamma(-{\epsilon_{+}\over 2\omega}+{1\over
2})}\label{b20}\\
  &  &  \nonumber\\
{1\over \sqrt{\lambda}}\frac{\Gamma(-{\epsilon_{-}\over
2\lambda}+{1\over 2})}{\Gamma(-{\epsilon_{-}\over 2\lambda}+1)} & = &
{1\over \sqrt{\omega}}\frac{\Gamma(-{\epsilon_{-}\over
2\omega}+{1\over 2})}{\Gamma(-{\epsilon_{+}\over
2\omega}+1)}\label{b21}
\end{eqnarray}
It is clear now that, as $\lambda\rightarrow\infty$, eqs. (\ref{b20})
and (\ref{b21}) give respectively
\begin{eqnarray}
\epsilon_{+,n} & = & 2\omega n\nonumber\\
  &  &  \qquad\hspace{.5in} n=0,1,2,\cdots\nonumber\\
\epsilon_{-,n} & = & 2\omega(n+1)\label{b22}
\end{eqnarray}
The corresponding wave functions are given by
\begin{eqnarray}
\psi_{+,n}(x) & = & B_{+,n}\,e^{-{1\over 2}\omega
x^{2}}\,H_{2n}(\sqrt{\omega}\, x)\nonumber\\
\psi_{-,n}(x) & = & B_{-,n}\,e^{-{1\over 2}\omega
x^{2}}\,H_{2n+1}(\sqrt{\omega}\, x)\label{b23}
\end{eqnarray}
These are, of course, the same energy levels and wave functions as
obtained in eqs. (\ref{b16}) and (\ref{b17}) respectively showing
again that
supersymmetry is manifest. Furthermore, this shows that this
conclusion is independent of the regularization used as long as the
regularization preserves supersymmetry which can be achieved by
properly regularizing the superpotential.

\section{Oscillator with ${1\over x^{2}}$ Potential}

In the last section, we showed that, in the presence of one kind of
singularity, namely, a boundary, supersymmetry is unbroken. In what
follows, we will study another class of supersymmetric models, namely,
the supersymmetric oscillator with a ${1\over x^{2}}$ potential, where
there is a genuine singularity in the potential not necessarily
arising  from a boundary. A naive analysis of this model [11] also
shows that supersymmetry is broken by such a singular potential (for
certain parameter ranges). However, this conclusion can be understood,
again, as a consequence of regularizing the potential
which, as we have seen before, does not respect supersymmetry. In
stead, we will show through a careful analysis that,
when the superpotential is regularized, supersymmetry is manifest in
this model as well (with a lot of interesting features).  
In this section, however, we will systematically analyze only the quantum
mechanical system corresponding to an oscillator in the presence of a
${1\over x^{2}}$ potential (postponing the discussion of the
supersymmetric  case to the next section). This system has been
analyzed  by several people [14-16]  and
the most complete analysis appears to be in ref. [16]. However, we feel
that, while the energy levels derived in [16] are correct, the wave
functions are not (namely, the extensions of the solutions from the
positive to the negative axis are incomplete and the wave functions,
of course, become quite crucial  when one
wants to extend the analysis to a supersymmetric system) and,
consequently, we  present a careful analysis of this system
regularizing the singular potential in a systematic manner. With the
supersymmetric system in mind (to follow in the next section), we
write the potential for the system as (with $\hbar=m=\omega=1$)
\begin{equation}
V(x) = {1\over 2}\left[\frac{g(g+1)}{x^{2}} + x^{2} - 2g +
1\right]\label{c1}
\end{equation}
Consequently, the Schr\"{o}dinger equation that we want to study has
the form
\begin{equation}
\left[\frac{d^{2}}{dx^{2}} - \frac{g(g+1)}{x^{2}} - x^{2} + (2\epsilon
+2g -1)\right]\psi(x) = 0\label{c2}
\end{equation}

The singular potential is repulsive for $g>0$ or $g<-1$ while it is
attractive for $-1<g<0$. Furthermore, for ease of comparison,
let us make the identifications with the notations of ref. [16] (note
that our energies are shifted since we have in mind the supersymmetric system
to study later)
\begin{eqnarray}
\lambda & = & {1\over 2}(\alpha^{2}-{1\over 4}) = {g(g+1)\over
2}\nonumber\\
E & = & \epsilon + g - {1\over 2}\label{c3}
\end{eqnarray}
It is also worth noting here that the Schr\"{o}dinger equation in
(\ref{c2}) is invariant under
\begin{eqnarray}
g & \leftrightarrow & -(g+1)\nonumber\\
\epsilon & \leftrightarrow & \epsilon + 2g + 1\label{c4}
\end{eqnarray}
This symmetry, of course, would also be reflected in the solutions.
Furthermore, 
the fixed point of this symmetry, namely, $g=-{1\over 2}$ separates
the two branches (namely, for every value of $\lambda$ there exist two
distinct values of $g$ corresponding to two distinct branches
separated at the branch point) in the parameter space.

\subsection*{Regularized Potential}

The Schr\"{o}dinger equation in (\ref{c2}) can be solved quite easily
for $x>0$ as was also done in [16]. However, to determine correctly
how this wavefunction should be extended to the negative axis, it is
more suitable to regularize the potential near the origin and study
the problem carefully. Let us consider a potential of the form
\begin{equation}
V(x) = \left\{\begin{array}{lll}
{1\over 2}\left[\frac{g(g+1)}{x^{2}}+x^{2}-2g+1\right] & {\rm for} &
|x|>R\\
 & & \\
{1\over 2}\left[\frac{g(g+1)}{R^{2}}+R^{2}-2g+1\right] & {\rm for} &
|x|<R
\end{array}\right.\label{c5}
\end{equation}
Namely, we have regularized the potential in a continuous manner
preserving the symmetry in eq. (\ref{c4}) with the understanding that
the regularization parameter $R\rightarrow 0$ at the end. With this
regularization, the Schr\"{o}dinger equation has to be analyzed in
three distinct regions. However, since the potential has reflection
symmetry, we need to analyze the solutions only in the regions
$-R<x<R$ and $x>R$.

The potential is a constant in the region $-R<x<R$ and hence the
Schr\"{o}dinger equation is quite simple here. The solutions can be
classified into even and odd ones and take the forms
\begin{eqnarray}
\psi^{(II)even}(x) & = & A(R) \cosh \kappa x\nonumber\\
\psi^{(II)odd}(x)  & = & B(R) \sinh \kappa x\label{c6}
\end{eqnarray}
where we have defined
\begin{equation}
\kappa = \sqrt{{g(g+1)\over R^{2}}+R^{2}-(2\epsilon + 2g - 1)}\approx
{\sqrt{g(g+1)}\over R}\label{c7}
\end{equation}
Since $R$ is small (and we are to take the vanishing limit at the
end), the last equality holds only if $g\neq 0\,{\rm or}\, -1$ which we will
assume. The special values of $g$ corresponding to the absence of a
singular potential have to be treated separately and we will come back
to this at the end of this section. We note here that the
normalization constants, $A$ and $B$, can, in principle depend on the
regularization parameter which we have allowed for in writing down the
form of the solutions in eq. (\ref{c6}).

The potential is much more complicated in the region $x>R$. However,
if we make the definitions
\begin{equation}
\xi = x^{2},\qquad \psi^{(I)} = e^{-{1\over 2}\xi}\,\xi^{{s\over
2}}\,u(\xi)\label{c8}
\end{equation}
where 
\begin{equation}
s = (g+1)\qquad {\rm or}\; -g\label{c9}
\end{equation}
the Schr\"{o}dinger equation of (\ref{c2}) takes the form
\begin{equation}
\xi\frac{d^{2}u}{d\xi^{2}} + (s+{1\over 2}-\xi)\frac{du}{d\xi}+{1\over
2}(\epsilon +g-s-1)u = 0\label{c10}
\end{equation}

An equation of the form
\begin{equation}
\xi\frac{d^{2}u}{d\xi^{2}} + (b-\xi)\frac{du}{d\xi}- a u =
0\label{c11}
\end{equation}
is known as the confluent hypergeometric equation [17] and the only
solution of this equation which is damped for large values of the
coordinate has the form
\begin{equation}
U(a,b,\xi) = \frac{\Gamma(1-b)}{\Gamma(1+a-b)}M(a,b,\xi) +
\frac{\Gamma(b-1)}{\Gamma(a)}\xi^{1-b}M(1+a-b,2-b,\xi)\label{c12}
\end{equation}
Here $M(a,b,\xi)$ are known as the confluent hypergeometric functions
which have the series expansion
\[
M(a,b,\xi) = 1 + {a\over b}\xi + {a(a+1)\over b(b+1)}{\xi^{2}\over 2!}
+ \cdots
\]
and satisfy
\begin{equation}
M(a,b,\xi) \stackrel{\xi\rightarrow 0}{\longrightarrow} 1\label{c13}
\end{equation}

It is clear now that eq. (\ref{c10}) simply is the confluent
hypergeometric equation and the asymptotically damped physical
solutions are nothing other than the $U(a,b,\xi)$ functions with
appropriate parameters. Since $s$ takes two possible values (see
eq. (\ref{c9})), it would seem that there would be two independent
solutions of eq. (\ref{c10}). However, it can be easily checked that
the two solutions corresponding to the two values of $s$ are really
proportional to each other and not independent. Thus, we can write the
general solution of the Schr\"{o}dinger equation, for $x>0$, as 
\begin{eqnarray}
\psi^{(I)}(x) & = & C(R)\,e^{-{1\over
2}x^{2}}\label{c14}\\
 &\times  &\left[\frac{\Gamma(-g-{1\over 2})}{\Gamma({1\over
2}-g-{\epsilon\over 2})}x^{g+1}\,M(1-{\epsilon\over 2},g+{3\over
2},x^{2}) + \frac{\Gamma(g+{1\over 2})}{\Gamma(1-{\epsilon\over
2})}x^{-g}\,M({1\over 2}-g-{\epsilon\over 2},-g+{1\over
2},x^{2})\right]\nonumber
\end{eqnarray}
Once again, we have allowed for a dependence of the normalization
constant, $C$, on the regularization parameter, $R$. However, for a
nontrivial solution to exist, we require that
\[
C(R)\stackrel{R\rightarrow 0}{\longrightarrow} C\neq 0
\]

So far, we have the general solutions, in the two regions, where
energy  is not quantized and 
which should arise from the matching conditions. Furthermore, we have
not bothered to evaluate the solution in the region $x<-R$ which
clearly would be the same as in the region $x>R$. However, the
matching conditions would determine  how we should extend the solutions
in the region $x>R$ to the region $x<-R$. Therefore, let us now
examine the matching conditions systematically since there are two
possible  cases.

\noindent$(i)$ {\bf\underline{Even Solution}}

We can match the even solution of the region $-R<x<R$ and its
derivative with those of the region $x>R$ at $x=R$. Taking the ratio
and remembering that $R$ is small (which is to be taken to zero at the
end), we obtain to the leading order in $R$
\begin{equation}
\sqrt{g(g+1)}\tanh \sqrt{g(g+1)} =\frac{(g+1){\Gamma(-g-{1\over
2})\over \Gamma({1\over 2}-g-{\epsilon\over
2})}R^{g+1}-g{\Gamma(g+{1\over 2})\over \Gamma(1-{\epsilon\over
2})}R^{-g}}{{\Gamma(-g-{1\over
2})\over \Gamma({1\over 2}-g-{\epsilon\over
2})}R^{g+1}+{\Gamma(g+{1\over 2})\over \Gamma(1-{\epsilon\over
2})}R^{-g}}\label{c15}
\end{equation}
Since the left hand side is independent of $R$, for consistency, the
right hand side must also be and this can happen in two different
ways.

First, for $g>-{1\over 2}$, it is clear that relation (\ref{c15}) can
be satisfied if (we assume from now on that $n=0,1,2,\cdots$.)
\begin{eqnarray}
1-{\epsilon\over 2} & = & -n + f_{1}(g)R^{2g+1}\nonumber\\
{\rm or,}\qquad \epsilon_{n} & = & 2(n+1) -
2f_{1}(g)R^{2g+1}\label{c16}
\end{eqnarray}
with a suitable choice of $f_{1}(g)$.

On the other hand, for $g<-{1\over 2}$, if
\begin{eqnarray}
{1\over 2}-g-{\epsilon\over 2} & = & -n + f_{2}(g)R^{-2g-1}\nonumber\\
{\rm or,}\qquad \epsilon_{n} & = & (2n-2g+1) -
2f_{2}(g)R^{-2g-1}\label{c17}
\end{eqnarray}
relation (\ref{c15}) can be satisfied with a suitable choice of
$f_{2}(g)$. It is clear that the two possible branches of the solution
simply reflect the symmetry in eq. (\ref{c4}).

This analysis shows that when the regularization is removed (namely,
$R\rightarrow 0$), we have an even extension of the solution of the
forms
\begin{equation}
g>-{1\over 2}:\qquad \epsilon_{n} = 2(n+1)\label{c18}
\end{equation}
with
\begin{equation}
\psi_{n}(x) = C_{n}\,\frac{\Gamma(-g-{1\over 2})}{\Gamma(-g-{1\over
2}-n)}e^{-{1\over 2}x^{2}}\,M(-n,g+{3\over
2},x^{2})\left\{\begin{array}{cll}
                x^{g+1}& {\rm for} & x>0\\
                |x|^{g+1}& {\rm for} & x<0
               \end{array}\right.\label{c19}
\end{equation}

\begin{equation}
g<-{1\over 2}:\qquad \epsilon_{n} = 2n -2g + 1\label{c20}
\end{equation}
with
\begin{equation}
\psi_{n}(x) = C_{n}\,\frac{\Gamma(g+{1\over 2})}{\Gamma(g+{1\over
2}-n)}e^{-{1\over 2}x^{2}}\,M(-n,-g+{1\over
2},x^{2})\left\{\begin{array}{cll}
                x^{-g}& {\rm for} & x>0\\
                |x|^{-g}& {\rm for} & x<0
               \end{array}\right.\label{c21}
\end{equation}

\noindent $(ii)$ {\bf\underline{Odd Solution}}

We can also match the odd solution of the region $-R<x<R$ and its
derivative with those of the region $x>R$  at $x=R$ and taking the
ratio, we obtain to leading order
\begin{equation}
\sqrt{g(g+1)}\coth \sqrt{g(g+1)} =\frac{(g+1){\Gamma(-g-{1\over
2})\over \Gamma({1\over 2}-g-{\epsilon\over
2})}R^{g+1}-g{\Gamma(g+{1\over 2})\over \Gamma(1-{\epsilon\over
2})}R^{-g}}{{\Gamma(-g-{1\over
2})\over \Gamma({1\over 2}-g-{\epsilon\over
2})}R^{g+1}+{\Gamma(g+{1\over 2})\over \Gamma(1-{\epsilon\over
2})}R^{-g}}\label{c22}
\end{equation}
Clearly, the analysis following from eq. (\ref{c15}) goes through
identically so that we  conclude that in the limit $R\rightarrow
0$, we have an odd extension of the solution of the forms
\begin{equation}
g>-{1\over 2}:\qquad \epsilon_{n} = 2(n+1)\label{c23}
\end{equation}
with
\begin{equation}
\psi_{n}(x) = C_{n}\,\frac{\Gamma(-g-{1\over 2})}{\Gamma(-g-{1\over
2}-n)}e^{-{1\over 2}x^{2}}\,M(-n,g+{3\over
2},x^{2})\left\{\begin{array}{cll}
                x^{g+1}& {\rm for} & x>0\\
                -|x|^{g+1}& {\rm for} & x<0
               \end{array}\right.\label{c24}
\end{equation}

\begin{equation}
g<-{1\over 2}:\qquad \epsilon_{n} = 2n -2g + 1\label{c25}
\end{equation}
with
\begin{equation}
\psi_{n}(x) = C_{n}\,\frac{\Gamma(g+{1\over 2})}{\Gamma(g+{1\over
2}-n)}e^{-{1\over 2}x^{2}}\,M(-n,-g+{1\over
2},x^{2})\left\{\begin{array}{cll}
                x^{-g}& {\rm for} & x>0\\
                -|x|^{-g}& {\rm for} & x<0
               \end{array}\right.\label{c26}
\end{equation}

\subsection*{Understanding  of the Result}

The conclusion following from this analysis, therefore, is that every
energy level of this system is doubly degenerate. Both even and odd
extensions of the solution are possible for every value of the energy
level. The energy levels, as given in eqs. (\ref{c18}) and (\ref{c20})
(or, alternately, (\ref{c23}) and (\ref{c25})) are, of course,
identical to those obtained in [16]. The crucial difference is in the
structure of the wave functions, namely, that both even and odd
extensions of the solution are possible for every value of the
energy (Incidentally, the solutions we have obtained in terms of
confluent hypergeometric functions also coincide with generalized
Laguerre polynomials as was obtained in ref. [16].). It is crucial,
therefore, to ask if such a conclusion is
physically plausible. To understand this question, let us recapitulate
the results from a simple quantum mechanical model which is well
studied. Namely, let us look at a particle moving in a potential of
the form
\[
V(x) = \left\{\begin{array}{cll}
              \gamma \delta(x) & {\rm for} & |x|<a\\
              \infty & {\rm for} & |x|>a
              \end{array}\right.
\]
It is well known that the solutions of this system can be classified
into even and odd ones with energy levels ($\hbar=m=1$)
\begin{eqnarray*}
E_{n}^{even} & = & \frac{n^{2}\pi^{2}}{2(a+{1\over
\gamma})^{2}}\nonumber\\
E_{n}^{odd} & = & \frac{n^{2}\pi^{2}}{2a^{2}}
\end{eqnarray*}
The even and the odd solutions, of course, have distinct energy values for
any finite strength of the delta potential. However, when
$\gamma\rightarrow\infty$, both the even and the odd solutions become
degenerate in energy. Namely, a delta potential with an infinite
strength leads to a double degeneracy of every energy level corresponding
to both even and odd solutions. The connection of this example with
the problem we are studying is intuitively clear. Namely, we can think
of
\[
{g(g+1)\over x^{2}} = \lim_{\eta\rightarrow 0}\,{g(g+1)\over
x^{2}+\eta^{2}} = \lim_{\eta\rightarrow 0}\left({\pi g(g+1)\over
\eta}\right)\,\left({1\over \pi}{\eta\over x^{2}+\eta^{2}}\right)
\]
It is clear that for $g\neq 0\,{\rm or}\,-1$, the singular ${1\over x^{2}}$
potential behaves like a delta potential with an infinite strength and
it is quite natural, therefore, that this system has both even and odd
solutions degenerate in energy.

It is also clear from this analysis that it is meaningless to take the
$g=0\,{\rm or}\,-1$ limit from the results obtained so far simply
because  the characters
of the two problems are quite different. As we have argued, for any
finite value of $g$ not coinciding with those special values, the
potential behaves, at the origin, like a delta potential of infinite
strength while for the special values, there is no such potential. The
two cases are related in a drastically discontinuous manner. As a
result, one cannot treat the ${g(g+1)\over x^{2}}$ as a perturbation
and obtain the full, correct solution simply because there is nothing
perturbative (small) about this potential for any \lq\lq nontrivial''
value of $g$. Another way of saying this is to re-emphasize what we
have already observed following eq. (\ref{c7}), namely, the character
of $\kappa$ and, therefore, the matching conditions change depending
on whether or not $g$ differs from the special values $0,-1$.

To see how the standard results of the harmonic oscillator would
emerge from this analysis, let us work out  the case only for $g=0$. In
this case, the matching conditions for the even solution would lead to
the relation  for the ratios
\begin{equation}  
-(2\epsilon-1)R = \frac{{\Gamma(-{1\over 2})\over \Gamma({1\over
 2}-{\epsilon\over 2})}-(2\epsilon-1)R\,{\Gamma({1\over 2})\over
 \Gamma(1-{\epsilon\over 2})}}{R\,{\Gamma(-{1\over 2})\over \Gamma({1\over
 2}-{\epsilon\over 2})}+{\Gamma({1\over 2})\over
 \Gamma(1-{\epsilon\over 2})}}\label{c27}
\end{equation}
which can be satisfied only if
\begin{equation}
{1\over 2}-{\epsilon\over 2} = -n,\qquad {\rm or,}\;\epsilon_{n} =
(2n+1)\label{c28}
\end{equation}
We recognize these to be the even levels of the oscillator (remember
the shifted zero point energy of the system in eq. (\ref{c1}) for
$g=0$). Similarly, matching the odd solution and its derivative leads
to
\begin{equation}
R=\frac{R\,{\Gamma(-{1\over 2})\over \Gamma({1\over
 2}-{\epsilon\over 2})}+{\Gamma({1\over 2})\over
 \Gamma(1-{\epsilon\over 2})}}{{\Gamma(-{1\over 2})\over \Gamma({1\over
 2}-{\epsilon\over 2})}-(2\epsilon-1)R\,{\Gamma({1\over 2})\over
 \Gamma(1-{\epsilon\over 2})}}\label{c29}
\end{equation}
which can be satisfied only if
\begin{equation}
1-{\epsilon\over 2} = -n,\qquad {\rm or,}\;\epsilon_{n} =
2(n+1)\label{c30}
\end{equation}
These are, of course, the odd energy levels of the harmonic
oscillator. (The corresponding wavefunctions are the even and odd
Hermite polynomials respectively.) There is no longer any degeneracy
of these levels. In
other words, the matching conditions change drastically and so do the
solutions of the problem depending on whether or not $g$ equals one of
the special values $0,-1$ and the results for the special values
cannot be (and, in fact, should not be) obtained from the general
result in a limiting manner.

\section{Supersymmetric Oscillator with ${1\over x^{2}}$ Potential}

In this section, we will analyze the supersymmetric version of the
case studied in the last section. Let us consider a superpotential of
the form [11]
\begin{equation}
W(x) = {g\over x} - x\label{d1}
\end{equation}
so that the pair of supersymmetric potentials would have the form
\begin{eqnarray}
V_{+}(x) & = & {1\over 2}\left[{g(g-1)\over
x^{2}}+x^{2}-2g-1\right]\nonumber\\
V_{-}(x) & = & {1\over 2}\left[{g(g+1)\over
x^{2}}+x^{2}-2g+1\right]\label{d2}
\end{eqnarray}
To analyze this problem, we should, of course, regularize the
superpotential. However, even before introducing the regularization,
let us observe some general features associated with this system, namely,
that this supersymmetric system would have a ground state satisfying
\begin{eqnarray}
Q\psi_{0}(x) & = & 0\nonumber\\
{\rm or,}\;\psi_{0}(x) & = & \left({x\over a}\right)^{g}\,e^{-{1\over
2}(x^{2}-a^{2})}\,\psi_{0}(a)\label{d3}
\end{eqnarray}
which would be damped for asymptotically large values of the
coordinate. On the other hand, from the behavior of this wave function
for small values of $x$, it is clear that normalizability of the wave
function would require that $g>-{1\over 2}$. We also note that since
the supersymmetric system involves two Hamiltonians with different $g$
dependence, the symmetry observed in the previous section, namely,
$g\leftrightarrow -(g+1)$ (or, alternately, $g\leftrightarrow -(g-1)$)
cannot be a symmetry of the whole system (Another way of saying this
is to note that the superpotential has no such symmetry. We will
come back to  this question later in this section.). Furthermore, we
cannot naively take over the results from the previous section since,
as we have seen earlier, regularizing the potential directly may not
respect supersymmetry.

Therefore, to study this problem systematically, we regularize the
superpotential  as
\begin{equation}
W(x) = \theta(x-|R|)\left({g\over x}-x\right) +
\theta(R-|x|)\left({g\over R}-R\right)\,{x\over R}\label{d4}
\end{equation}
Here, as before, $R$ is the regularization parameter which should be
taken to zero at the end and we have regularized the superpotential
such that it is continuous across the boundary. This has the nice
feature that there are no delta potential terms in the potentials. In
fact, the regularized superpotential leads to the pair of
supersymmetric potentials of the forms
\begin{eqnarray}
V_{+}(x) & = &\!\!\! {1\over 2}\left[\theta(x-|R|)\left({g(g-1)\over
x^{2}}+x^{2}-2g-1\right) + \theta(R-|x|)\left(\left({g\over
R^{2}}-1\right)^{2}x^{2}+\left({g\over
R^{2}}-1\right)\right)\right]\nonumber\\
V_{-}(x) & = &\!\!\! {1\over 2}\left[\theta(x-|R|)\left({g(g+1)\over
x^{2}}+x^{2}-2g+1\right) + \theta(R-|x|)\left(\!\!\left({g\over
R^{2}}-1\right)^{2}x^{2}-\left({g\over
R^{2}}-1\right)\!\!\right)\!\right]\label{d5}
\end{eqnarray}
It is worth noting here that for $g=0$, the singular potential
at the origin is not present in both the Hamiltonians (namely, it is
truly not there). On the other hand, for $g=1\,{\rm or}\,-1$, although
the singular potential disappears from only one of the Hamiltonians,
the complete system remembers about it through supersymmetry as is
clear from the structure of the regularized potentials.

\subsection*{Spectrum of $H_{+}$}

Let us now analyze the spectrum of the two different Hamiltonians
systematically. First, let us note that, for the Hamiltonian $H_{+}$,
the potential is that of a harmonic oscillator in the region
$-R<x<R$. We can write down the even and the odd solutions in this
region as [17] (We will assume throughout that $g\neq 0$)
\begin{eqnarray}
\psi_{+}^{(II)even}(x) & = & A_{+}(R)\,e^{-{1\over 2}\left({g\over
R^{2}}-1\right)x^{2}}\,M(-{\epsilon_{+}R^{2}\over g-R^{2}}+{1\over
2},{1\over 2},({g\over R^{2}}-1)x^{2})\nonumber\\
\psi_{+}^{(II)odd}(x) & = & B_{+}(R)\,x\,e^{-{1\over 2}\left({g\over
R^{2}}-1\right)x^{2}}\,M(-{\epsilon_{+}R^{2}\over g-R^{2}}+1,{3\over
2},({g\over R^{2}}-1)x^{2})\label{d6}
\end{eqnarray}

The solution in the region $x>R$ can also be obtained from an
analysis as given in the earlier section and leads to
\begin{eqnarray}
\psi_{+}^{(I)}(x) & = & C_{+}(R)\,e^{-{1\over 2}x^{2}}\label{d7}\\
  &\times  &\left[\frac{\Gamma(-g+{1\over 2})}{\Gamma({1\over
  2}-g-{\epsilon_{+}\over 2})}x^{g}M(-{\epsilon_{+}\over 2},g+{1\over
  2},x^{2}) +\frac{\Gamma(g-{1\over 2})}{\Gamma(-{\epsilon_{+}\over
  2})}x^{-g+1} M({1\over 2}-g-{\epsilon_{+}\over 2},-g+{3\over
  2},x^{2})\right]\nonumber
\end{eqnarray}
Once again, we can match the solutions in the two regions and their
derivatives across $x=R$ to obtain the relevant quantization
conditions. As before, there are two possibilities.

\noindent $(i)$ {\bf\underline{Even Solution}}

If we match the even solution in eq. (\ref{d6}) and its derivative
with those of eq. (\ref{d7}) at $x=R$ and take the ratio, we obtain  to leading
order in $R$ (remember $R$ is small and is to be taken to zero at the
end),
\begin{eqnarray}
1 & = & \frac{{\Gamma(-g+{1\over 2})\over \Gamma({1\over
2}-g-{\epsilon_{+}\over 2})}R^{g}-{1\over g}(g-1){\Gamma(g-{1\over
2})\over \Gamma(-{\epsilon_{+}\over 2})}R^{-g+1}}{{\Gamma(-g+{1\over
2})\over \Gamma({1\over
2}-g-{\epsilon_{+}\over 2})}R^{g}+{\Gamma(g-{1\over
2})\over \Gamma(-{\epsilon_{+}\over 2})}R^{-g+1}}\nonumber\\
 &  &  \nonumber\\
 & = & 1-{(2g-1)\over g}\frac{{\Gamma(g-{1\over 2})\over
\Gamma(-{\epsilon_{+}\over 2})}R^{-g+1}}{{\Gamma(-g+{1\over
2})\over \Gamma({1\over
2}-g-{\epsilon_{+}\over 2})}R^{g}+{\Gamma(g-{1\over
2})\over \Gamma(-{\epsilon_{+}\over 2})}R^{-g+1}}\label{d8}
\end{eqnarray}
Clearly, this can be satisfied only if ($n=0,1,2,\cdots$)
\begin{equation}
-{\epsilon_{+}\over 2} = -n,\qquad {\rm or,}\,\epsilon_{+,n} =
 2n\label{d9}
\end{equation}
Thus, we obtain that there exists an even extension of the solution of
the form (when the regularization is removed)
\begin{equation}
\psi_{+,n}(x) = C_{+,n}\frac{\Gamma(-g+{1\over 2})}{\Gamma(-g+{1\over
2}-n)}e^{-{1\over 2}x^{2}}\,M(-n,g+{1\over
2},x^{2})\left\{\begin{array}{cll}
                x^{g}& {\rm for} & x>0\\
                |x|^{g}& {\rm for} & x<0
               \end{array}\right.\label{d10}
\end{equation}
with the energy levels given by
\begin{equation}
\epsilon_{+,n} = 2n\label{d11}
\end{equation}
Furthermore, we note from equation (\ref{d10}) that normalizability of
the solution restricts that this solution is physical only for
$g>-{1\over 2}$.

\noindent (ii) {\bf\underline{Odd Solution}}

We can also match the odd solution in eq. (\ref{d6}) and its
derivative with those of  eq. (\ref{d7}) and the ratio leads to the
appropriate quantization conditions. Without going into technical
details, let us simply note here that there are two possibilities in
this case. For $g>{1\over 2}$, we have in the limit $R\rightarrow 0$
\begin{equation}
\epsilon_{+,n} = 2n \label{d12}
\end{equation}
with
\begin{equation}
\psi_{+,n}(x) = C_{+,n}\frac{\Gamma(-g+{1\over 2})}{\Gamma(-g+{1\over
2}-n)}e^{-{1\over 2}x^{2}}\,M(-n,g+{1\over
2},x^{2})\left\{\begin{array}{cll}
                x^{g}& {\rm for} & x>0\\
                -|x|^{g}& {\rm for} & x<0
               \end{array}\right.\label{d13}
\end{equation}
On the other hand, for $g<{1\over 2}$, 
\begin{equation}
\epsilon_{+,n} = 2n-2g+1\label{d14}
\end{equation}
with
\begin{equation}
\psi_{+,n}(x) = C_{+,n}\frac{\Gamma(g-{1\over 2})}{\Gamma(g-{1\over
2}-n)}e^{-{1\over 2}x^{2}}\,M(-n,-g+{3\over
2},x^{2})\left\{\begin{array}{cll}
                x^{-g+1}& {\rm for} & x>0\\
                -|x|^{-g+1}& {\rm for} & x<0
               \end{array}\right.\label{d15}
\end{equation}
This completes the analysis of the spectrum for $H_{+}$. We see that
it consists of a set of  even solutions with vanishing energy as we would
expect from supersymmetry, but it also contains additional physical
solutions. 

\subsection*{Spectrum of $H_{-}$}

We can similarly analyze the spectrum for the supersymmetric partner
Hamiltonian $H_{-}$. The even and the odd solutions in the region
$-R<x<R$  have the forms
\begin{eqnarray}
\psi_{-}^{(II)even}(x) & = & A_{-}(R)\,e^{-{1\over 2}\left({g\over
R^{2}}-1\right)x^{2}}\,M(-{\epsilon_{-}R^{2}\over g-R^{2}},{1\over 2},
({g\over R^{2}}-1)x^{2})\nonumber\\
\psi_{-}^{(II)odd}(x) & = & B_{-}(R)\,x\,e^{-{1\over 2}\left({g\over
R^{2}}-1\right)x^{2}}\,M(-{\epsilon_{-}R^{2}\over g-R^{2}}+{1\over 2},{3\over
2},({g\over R^{2}}-1)x^{2})\label{d16}
\end{eqnarray}
while the solution in the region $x>R$ now takes the form
\begin{eqnarray}
\psi_{-}^{(I)}(x) & = & C_{-}(R)\,e^{-{1\over 2}x^{2}}\label{d17}\\
  &\times  &\!\!\!\left[\frac{\Gamma(-g-{1\over 2})}{\Gamma({1\over
  2}-g-{\epsilon_{-}\over 2})}x^{g+1}M(1-{\epsilon_{-}\over 2},g+{3\over
  2},x^{2}) +\frac{\Gamma(g+{1\over 2})}{\Gamma(1-{\epsilon_{-}\over
  2})}x^{-g} M({1\over 2}-g-{\epsilon_{-}\over 2},-g+{1\over
  2},x^{2})\right]\nonumber
\end{eqnarray}

The matching of the solutions and their derivatives can now be done in
the standard manner and we simply give the results in the limit
$R\rightarrow 0$ here. There exists a normalizable even solution only
when $g<{1\over 2}$ of the form
\begin{equation}
\psi_{-,n}(x) = C_{-,n}\frac{\Gamma(g+{1\over 2})}{\Gamma(g+{1\over
2}-n)}e^{-{1\over 2}x^{2}}\,M(-n,-g+{1\over
2},x^{2})\left\{\begin{array}{cll}
                x^{-g}& {\rm for} & x>0\\
                |x|^{-g}& {\rm for} & x<0
               \end{array}\right.\label{d18}
\end{equation}
with
\begin{equation}
\epsilon_{-,n} = 2n-2g+1\label{d19}
\end{equation}
On the other hand, there are two possible odd solutions. For
$g>-{1\over 2}$, we have
\begin{equation}
\psi_{-,n}(x) = C_{-,n}\frac{\Gamma(-g-{1\over 2})}{\Gamma(-g-{1\over
2}-n)}e^{-{1\over 2}x^{2}}\,M(-n, g+{3\over
2},x^{2})\left\{\begin{array}{cll}
                x^{g+1}& {\rm for} & x>0\\
                -|x|^{g+1}& {\rm for} & x<0
               \end{array}\right.\label{d20}
\end{equation}
with
\begin{equation}
\epsilon_{-,n} = 2(n+1)\label{d21}
\end{equation}
while, for $g<-{1\over 2}$, we have
\begin{equation}
\psi_{-,n}(x) = C_{-,n}\frac{\Gamma(g+{1\over 2})}{\Gamma(g+{1\over
2}-n)}e^{-{1\over 2}x^{2}}\,M(-n,-g+{1\over
2},x^{2})\left\{\begin{array}{cll}
                x^{-g}& {\rm for} & x>0\\
                -|x|^{-g}& {\rm for} & x<0
               \end{array}\right.\label{d22}
\end{equation}
with
\begin{equation}
\epsilon_{-,n} = 2n-2g+1\label{d23}
\end{equation}
This completes the determination of the spectrum of the supersymmetric
pair of Hamiltonians, $H_{+}$ and $H_{-}$. We note that we have many
more states than would be needed from the point of view of
supersymmetry and, therefore, it is crucial to understand the
solutions in a systematic manner which we do next.

\subsection*{Analysis of the Result}

The large number of solutions obtained is really very interesting and
to appreciate their presence, let us analyze their behavior in some
detail. First, let us note that (as we had pointed out in
eq. (\ref{d3}) and in the subsequent discussion) a supersymmetric
ground state would exist for our system only if $g>-{1\over
2}$. Indeed, there exists such a branch of the solutions in our
model, namely, the even solutions of $H_{+}$ and one of the two sets of
odd solutions of $H_{-}$. They are, in fact, degenerate in energy
except for the ground state which is an even state with vanishing
energy belonging to the spectrum of $H_{+}$. The solutions have the
relative odd parity appropriate to be superpartner states and, in
fact, it is not hard to check using the properties of the confluent
hypergeometric functions [17], that, for $g>-{1\over 2}$,
\begin{eqnarray}
Q^{\dagger}(g)\psi_{-,n}^{odd}(x) & = & {1\over \sqrt{2}}\left({d\over
dx}+{g\over x}-x\right) \psi_{-,n}^{odd}(x)\nonumber\\
 & = & -\sqrt{2}\,{C_{-,n}\over C_{+,n+1}}\,\psi_{+,n+1}^{even}(x)\label{d24}
\end{eqnarray}
as we would expect from states which are superpartners of each 
other. This shows that the system of solutions, at least, contains the
supersymmetric set that we are interested in.

This, therefore, raises the question about the roles of the other
solutions that we have found. Both $H_{+}$ and $H_{-}$ have solutions
on another common branch, namely, for $g<{1\over 2}$ and it is worth
investigating whether, they, too, define a supersymmetric set of
solutions. They are degenerate in energy (see eqs. (\ref{d14}) and
(\ref{d19})) and they have the correct relative odd parity structure
(solutions  of $H_{+}$
are odd while those for $H_{-}$ are even). However, it is easily
verified that
\begin{equation}
Q^{\dagger}(g)\psi_{-,n}^{even}(x) \neq
\psi_{+,n+1}^{odd}(x)\label{d25}
\end{equation}
even up to normalization constants. On the other hand, it can also be
checked equally easily that
\begin{eqnarray}
Q^{\dagger}(-g)\psi_{+,n}^{odd}(x) & = & {1\over
\sqrt{2}}\left({d\over dx}-{g\over x}-x\right)\psi_{+,n}^{odd}(x)\nonumber\\
  & = & -\sqrt{2} {C_{+,n}\over C_{-,n+1}}\psi_{-,n+1}^{even}(x)\label{d26}
\end{eqnarray}
Thus, there appears to be some sort of supersymmetry for these states,
but there is  no apparent state of vanishing energy. Furthermore,
there is, of
course, the third set of solutions of $H_{+}$ and $H_{-}$ which do not even
share the same branch.

The meaning of all these solutions becomes quite clear once we realize that the
supersymmetrization of a given Hamiltonian, in this case, is not
unique. There are, in fact, various possibilities. For example,

\noindent $(i)$ we can choose as the superpotential
\begin{equation}
W(x) = {g\over x}-x\label{d27}
\end{equation}
as we have done so that the supersymmetric pair potentials would be
\begin{eqnarray}
V_{+}(x) & = & {1\over 2}\left[{g(g-1)\over
x^{2}}+x^{2}-2g-1\right]\nonumber\\
V_{-}(x) & = & {1\over 2}\left[{g(g+1)\over
x^{2}}+x^{2}-2g+1\right]\label{d28}
\end{eqnarray}
Here, as we have already seen, a normalizable ground state would exist
only for $g>-{1\over 2}$ (The ground state belongs to $H_{+}$).

\noindent $(ii)$ If, on the other hand, we choose
\begin{equation}
W(x) = {g-1\over x}-x\label{d29}
\end{equation}
the supersymmetric pair of potentials become
\begin{eqnarray}
V_{+}(x) & = & {1\over 2}\left[{(g-1)(g-2)\over
x^{2}}+x^{2}-2g+1\right]\nonumber\\
V_{-}(x) & = & {1\over 2}\left[{g(g-1)\over
x^{2}}+x^{2}-2g+3\right]\label{d30}
\end{eqnarray}
where a normalizable ground state would exist only if $g>{1\over 2}$.

\noindent $(iii)$ Similarly, we could have chosen
\begin{equation}
W(x) = -{g+1\over x} - x\label{d31}
\end{equation}
which would lead to a pair of supersymmetric potentials
\begin{eqnarray}
V_{+}(x) & = & {1\over 2}\left[{(g+1)(g+2)\over
x^{2}}+x^{2}+2g+1\right]\nonumber\\
V_{-}(x) & = & {1\over 2}\left[{g(g+1)\over
x^{2}}+x^{2}+2g+3\right]\label{d32}
\end{eqnarray}
where a normalizable ground state would exist only for $g<-{1\over
2}$. 

In some sense, this is the generalization of the idea that a
spin ${1\over 2}$ particle, say for example, can belong to distinct
multiplets  of the form $(0,{1\over 2})$ or $({1\over 2},1)$. (As is
clear now, the transformation $g\leftrightarrow
-(g+1)$ or $g\leftrightarrow -(g-1)$ alluded to earlier would simply
take one supersymmetric system to another and cannot be a symmetry of
one supersymmetric system.)

\noindent $(iv)$ However, there is also the possibility of choosing
\begin{equation}
W(x) = -{g\over x} -x\label{d33}
\end{equation}
giving the supersymmetric potentials
\begin{eqnarray}
V_{+}(x) & = & {1\over 2}\left[{g(g+1)\over
x^{2}}+x^{2}+2g-1\right]\nonumber\\
V_{-}(x) & = & {1\over 2}\left[{g(g-1)\over
x^{2}}+x^{2}+2g+1\right]\label{d34}
\end{eqnarray}
Namely, $H_{+}$ and $H_{-}$ could  also have reversed their roles (This
basically determines  which potential contains the ground
state). In this case, a normalizable ground state would exist only for
$g<{1\over 2}$.

The meaning of all the solutions is clear now. First, the solutions
corresponding to the branch $g>-{1\over 2}$ define the supersymmetric
solutions corresponding to the supersymmetrization in
eq. (\ref{d27}). The solutions corresponding to the branch $g<{1\over
2}$ also define a supersymmetric set of solutions corresponding to
the supersymmetrization in eq. (\ref{d33}). (It is clear now why the
solutions were related to each other as in eq. (\ref{d26}).) The fact
that the spectrum contains a zero energy state, as we would expect in
a supersymmetric system, follows from the fact that the potentials in
(\ref{d34}) are shifted with respect to those in (\ref{d28}). Finally,
the other solutions that do not even share the same branch correspond
to the supersymmetrizations in eqs. (\ref{d29}) and
(\ref{d31}). Supersymmetry is manifest in spite of the singular
behavior of the potentials. It is also clear from this analysis that
distinct solutions really correspond to distinct
supersymmetrizations and in deriving conclusions regarding
supersymmetry,  one should be very
careful in identifying the correct solutions from among the whole set.

\section{Algebraic Solution}

The supersymmetric system studied in the previous section is a very
special system which can also be solved algebraically. Therefore, to
further understand the properties of this system, we present here a
short derivation of the algebraic solution of this system.

Let us consider the supersymmetric system of potentials in
eq. (\ref{d2}). Defining the supercharges as in (\ref{a3}), namely,
\begin{eqnarray}
Q(g) & = & {1\over \sqrt{2}}\left(-{d\over dx}+{g\over
x}-x\right)\nonumber\\
Q^{\dagger}(g) & = & {1\over \sqrt{2}}\left({d\over dx}+{g\over
x}-x\right)\label{e1}
\end{eqnarray}
we can write the supersymmetric pair of Hamiltonians also as
\begin{eqnarray}
H_{+}(g) & = & Q^{\dagger}(g)Q(g)\nonumber\\
H_{-}(g) & = & Q(g) Q^{\dagger}(g)\label{e2}
\end{eqnarray}
However, from the structure of the potentials in eq. (\ref{d2}), it is
clear that we can write 
\begin{equation}
H_{-}(g) = Q(g)Q^{\dagger}(g) = Q^{\dagger}(g+1)Q(g+1) + 2 =
H_{+}(g+1)+2\label{e3}
\end{equation}
Such a Hamiltonian, $H_{+}(g)$, (equivalently, the potential) is known
as shape invariant [18,12]. The ground state of $H_{+}(g)$ satisfies
\begin{eqnarray}
Q(g)\psi_{+,0}(g,x) & = & {1\over \sqrt{2}}\left({d\over dx}+{g\over
x}-x\right)\psi_{+,0}(x) = 0\nonumber\\
{\rm or,}\;\;\psi_{+,0}(g,x) & \approx & x^{g}\,e^{-{1\over
2}x^{2}}\label{e4}
\end{eqnarray}
This solution is normalizable and physical only if
\begin{equation}
g>-{1\over 2}\label{e5}
\end{equation}
Furthermore, the ground state energy of $H_{+}(g)$, in such a case, is
zero,
\begin{equation}
\epsilon_{+,0} = 0\label{e6}
\end{equation}
Since $H_{+}(g)$ and $H_{-}(g)$ are supersymmetric partners and
share all the energy levels except for the ground state of $H_{+}(g)$,
it is clear that the ground state energy of $H_{-}(g)$ would
correspond to the first excited state energy of $H_{+}(g)$ and from
the relations in eqs. (\ref{e3}) and (\ref{e6}), it follows,
therefore, that the energy of the first excited state of $H_{+}(g)$ is
\begin{equation}
\epsilon_{+,1} = 2\label{e7}
\end{equation}
Let us note here that the ground state wavefunction of $H_{-}(g)$
(from the structure in eq. (\ref{e3})) would follow to have the form
\begin{equation}
\psi_{+,0}(g+1,x) \approx x^{g+1}\,e^{-{1\over 2}x^{2}}\label{e8}
\end{equation}
and would be normalizable only if
\[
g>-{3\over 2}
\]
and this would automatically hold if eq. (\ref{e5}) is true.

We can, in fact, construct a sequence of Hamiltonians, in such a case, of
the form
\begin{eqnarray}
H^{(0)} & = & H_{+}(g)=Q^{\dagger}(g)Q(g)\nonumber\\
H^{(1)} & = &
H_{-}(g)=Q(g)Q^{\dagger}(g)=Q^{\dagger}(g+1)Q(g+1)+2\nonumber\\
H^{(2)} & = &
Q(g+1)Q^{\dagger}(g+1)+2=Q^{\dagger}(g+2)Q(g+2)+4\nonumber\\
\vdots &  & \nonumber\\
H^{(s)} & = & Q^{\dagger}(g+s)Q(g+s)+2s\label{e9}
\end{eqnarray}
From the structure of the Hamiltonians, it is clear that the ground
state of any Hamiltonian in the sequence would be  normalizable
if eq. (\ref{e5}) holds. Furthermore, every adjacent pair of
Hamiltonians, $H^{(s)}$ and $H^{(s-1)}$ would define a supersymmetric
pair of Hamiltonians and share all the energy levels except for the
ground state of $H^{(s-1)}$ which would have an energy eigenvalue
$2(s-1)$. The energy levels of our original pair of Hamiltonians,
$H_{+}(g)$ and $H_{-}(g)$ now follow from this to be
\begin{equation}
\epsilon_{+,n} = 2n,\qquad \epsilon_{-,n} = 2(n+1)\label{e10}
\end{equation}
These are of course, the energy levels obtained in our earlier
analysis (see eqs. (\ref{d9}) and (\ref{d21})).

The wave functions for the original Hamiltonians can also be obtained
from an analysis of the spectrum of this sequence of Hamiltonians. It
is easily seen that [19]
\begin{equation}
\psi_{+,n}(g,x) \propto  Q^{\dagger}(g)\psi_{+,n-1}(g+1,x)\label{e11}
\end{equation}
which can be iterated to give
\begin{equation}
\psi_{+,n}(g,x) \propto Q^{\dagger}(g)Q^{\dagger}(g+1)\cdots
Q^{\dagger}(g+n-1)\psi_{+,0}(g+n,x)\label{e12}
\end{equation}
And since, we know the ground state wave function, $\psi_{+,0}$ (see
eq. (\ref{e4})), all the wave functions for the higher states can be
explicitly constructed. Furthermore, once we know $\psi_{+,n}(g,x)$,
the superpartner states can also be obtained from the relation that (see
(\ref{a5}) and the discussion following)
\begin{equation}
\psi_{-,n}(g,x) \propto Q(g)\psi_{+,n}(g,x)\label{e13}
\end{equation}

It is worth asking if these algebraically constructed states also
coincide with the states we have constructed in eqs. (\ref{d10}) and
(\ref{d20}) respectively. It is easy to verify the relation that
\begin{equation}
\left({d\over dx}+{g\over x}-x\right)e^{-{1\over
2}x^{2}}\,x^{g+1}\,M(-n+1,g+{3\over 2},x^{2}) = (2g+1)\,e^{-{1\over
2}x^{2}}\,x^{g}\,M(-n,g+{1\over 2},x^{2})\label{e14}
\end{equation}
which is the appropriate recursion relation in eq. (\ref{e11}) for our
system. Furthermore, noting the form of the ground state wave function
in eq. (\ref{e4})
\begin{equation}
e^{-{1\over 2}x^{2}}\,x^{g} = e^{-{1\over
2}x^{2}}\,x^{g}\,M(0,g+{1\over 2},x^{2})\label{e15}
\end{equation}
and choosing an even function for the ground state, as we normally do,
(as far as we can
see, the algebraic method cannot determine {\it a priori} whether the
function should be even/odd), it is clear that all the states
associated with $H_{+}(g)$ would be even states and would coincide
with those in eq. (\ref{d10}). Once these are identified, the
superpartner states also follow to coincide with those in
eq. (\ref{d20}) and are odd. This concludes the algebraic construction
of the supersymmetric spectrum of the system and shows again that
supersymmetry is manifest.

\section{The Puzzle}

An interesting puzzle has been raised in the literature [11] in
connection  with the
supersymmetric  system of eq. (\ref{d2}) and in this section, we 
describe the resolution of the puzzle  from a
systematic analysis of the problem. Namely, let us look at the pair of
supersymmetric potentials in eq. (\ref{d2})
\begin{eqnarray*}
V_{+}(x) & = & {1\over 2}\left[{g(g-1)\over
x^{2}}+x^{2}-2g-1\right]\nonumber\\
V_{-}(x) & = & {1\over 2}\left[{g(g+1)\over
x^{2}}+x^{2}-2g+1\right]
\end{eqnarray*}
and note that, for $g=1$, the potentials take the form
\begin{eqnarray}
V_{+}(x) & = & {1\over 2}\left(x^{2}-3\right)\nonumber\\
V_{-}(x) & = & {1\over 2}\left({2\over x^{2}}+x^{2}-1\right)\label{f1}
\end{eqnarray}
As we have noted earlier, $g=1$ is not particularly a special point in
the parameter space (only $g=0$ is) and, consequently, our analysis
would continue to hold for this value of the coupling as
well. Furthermore, this value of the coupling is consistent with
$g>-{1\over 2}$ and, therefore, we would expect the spectrum to be
supersymmetric with the ground state energy of $H_{+}(g=1)$ 
vanishing as the analysis of the last two sections have
shown. On the other hand, if we look at the potential $V_{+}$ in
eq. (\ref{f1}), it is clear that it is a harmonic oscillator
potential with the zero point energy shifted by $-{3\over 2}$. It
would appear, therefore, that the ground state energy of $H_{+}(g=1)$
would be negative contrary to what our analysis has shown and
in violation of the theorems of supersymmetry (The supersymmetry
algebra tells us that the ground state energy in a supersymmetric
theory has to be positive semi-definite.). This, therefore, raises an
interesting puzzle and it has been discussed in the literature from
various points of view. However, the resolution of this puzzle is
really quite simple as we show next.

In concluding that the potential $V_{+}$ in eq. (\ref{f1}) would lead
to a negative energy, we have, of course, tacitly assumed the 
ground state wave function for the system to be the standard
oscillator wavefunction, namely,
\begin{equation}
\psi_{0}(x) \propto e^{-{1\over 2}x^{2}}\label{f2}
\end{equation}
which, as it should be, is an even function. On the other hand, from
our  analysis of the
last two sections (see eq. (\ref{d10}) or (\ref{e4})), we would
conclude that the ground state wave function for our supersymmetric
system would have the form
\begin{equation}
\psi_{+,0}(x) \propto e^{-{1\over 2}x^{2}}\,|x|\label{f3}
\end{equation}
It can also be explicitly checked that this indeed has a vanishing energy
consistent with our general analysis. Therefore, understanding the
question of the negative energy reduces to understanding the role of
the wave function in eq. (\ref{f2}). It is, of course, clear from the
explicit spectrum of the system  we have constructed that it is not
part of our supersymmetric system. It must, therefore, belong to a
different supersymmetric system since as we have seen earlier,
distinct supersymmetrizations lead to distinct forms of the wave
functions. 

In fact, let us choose
\begin{equation}
\widetilde{W}(x) = -{g-1\over x}-x\label{f4}
\end{equation}
which would lead to the pair of supersymmetric potentials of the forms
\begin{eqnarray}
\widetilde{V}_{+}(x) & = & {1\over 2}\left[{g(g-1)\over
x^{2}}+x^{2}+2g-3\right]\nonumber\\
\widetilde{V}_{-}(x) & = & {1\over 2}\left[{(g-1)(g-2)\over
x^{2}}+x^{2}+2g-1\right]\label{f5}
\end{eqnarray}
It is clear that this supersymmetric system would have a normalizable,
even  ground state of the form
\begin{equation}
\widetilde{\psi}_{0}(x) \propto |x|^{1-g}\,e^{-{1\over 2}x^{2}}\label{f6}
\end{equation}
which is normalizable for $g<{3\over 2}$. It is also clear that this
wave function coincides with (\ref{f2}) when $g=1$ (which is in its
range of validity). In other words, the wave function in
eq. (\ref{f2}) does not belong to the supersymmetric spectrum of
(\ref{d2}), rather, it is a legitimate ground state of the
supersymmetrization in eq. (\ref{f4}). For $g=1$ the potentials of (\ref{f5}),
of course, have the form
\begin{eqnarray}
\widetilde{V}_{+}(x) & = & {1\over 2}\left[x^{2}-1\right]\nonumber\\
\widetilde{V}_{-}(x) & = & {1\over 2}\left[x^{2}+1\right]\label{f7}
\end{eqnarray}
which is nothing other than the supersymmetric oscillator which, as  we
know, has a zero energy ground state.

Thus, there is really no puzzle and the theorems of supersymmetry are,
indeed, on firm footing. The crucial thing to learn from this
analysis is what we have emphasized earlier, namely, in dealing with
supersymmetry in such a system it is essential to identify the
appropriate solution relevant for a particular supersymmetric system because
every  distinct solution corresponds to a distinct supersymmetrization
of the system. 

\section{Conclusion}
In this paper, we have studied systematically two classes of
supersymmetric quantum  mechanical models - one consisting of a
singular boundary and the other with a singular potential. We have
shown that, contrary to the conventional understanding [10-12], supersymmetry
is manifest in these systems. In particular, for a system with a
singular potential such as ${1\over x^{2}}$, we have shown that the
solution of the Schr\"{o}dinger equation leads to several distinct
solutions corresponding to distinct supersymmetrizations of the
system. Consequently, it becomes quite important to identify the
appropriate wavefunctions when supersymmetric properties are being
investigated. We have also solved the system with a singular
potential algebraically using the ideas of shape invariance which
further corroborates the results of our direct analysis of the
Schr\"{o}dinger equation, namely, that supersymmetry is manifest in
this system. Furthermore, we have shown how a puzzle raised in the
literature finds a natural resolution when one does a careful analysis
restricting oneself to the appropriate branch of the supersymmetric
solutions. Finally, we would like to conclude by noting that
supersymmetry is known to be robust at short distances (high
energies). The singularities discussed in the quantum mechanical
models occur at short distances  and, therefore, it is intuitively quite
clear that they are unlikely to break supersymmetry. Our detailed,
systematic analysis only reinforces this.    

\section*{Acknowledgments}

This work was supported in part by the U.S. Dept. of
Energy Grant  DE-FG 02-91ER40685.

\end{document}